# A causal approach to first-order optical equivalency of inertial systems, by means of a beam-pointing test-experiment based on speed-induced deflection of light.

G. Sardin
gsardin@ub.edu

**Abstract**

Within the framework of test-experiments, an original pointing set-up based on speed-induced deflection of a light-beam and using a high-resolution opto-electronic array as a position detector, is proposed. The device would provide a new way to ratify the first order optical equivalency of inertial systems and to sound its causal groundwork. The type of test provided applies to any device using beams, but here it has been adapted for an up-dated check-up of Michelson's experiment and to bear out the interpretation derived from its null result. Rather than searching for a speed-induced phase-shift as Michelson's experiment was conceived for, the present set-up would instead detect any deviation from first-order equivalency through any speed-induced lateral-shift between the two split beams on recombining. In effect, any eventual slight shift would affect the beam path and hence the interferometric performances. Taking into account the conceptual transcendence of the deductions derived from this experiment and that it constitutes, through the invariance of its result, one of the most relevant historical experimental proofs considered to back the theory of special relativity, and in particular its first postulate on the exhaustive equivalency of inertial systems, it cannot be superfluous to carefully update its performances and the derived deductions. So, the presence as well as the absence of speed-induced beam deflection have been considered, and their respective implications analyzed and related to the null result of this archetypal interferometric experiment, which is still much adequate for a conceptual analysis. A causal approach to the first-order optical equivalency of inertial systems is advanced.

## I. Introduction

Here is proposed an original set-up to sound the origin of the equivalency of inertial systems. Its usefulness in regard to conceptual grounds is also discussed. In effect, we think that the experiment proposed should not be separated from its conceptual frame, otherwise the set-up utility may not be clearly perceived nor the novelty of its contribution. Let us point out that the most accurate experiments made to check the electromagnetic equivalency of inertial systems, and from which their formal equivalency has been deduced, are mainly of the interferometric type. Therefore, the beams path length is a crucial parameter for the interpretation of their experimental results. The set-up proposed has the purpose of checking this critical point, i.e. to determine the beam path and so its effective length. On conceptual grounds it should help developing a causal understanding of the optical equivalency of inertial systems. The proposed experiment has been correlated to Michelson's one in view that both are concerned with beams path and that the latter constitutes a widely known archetypal experiment, most appropriate for a phenomenological analysis, including those modern versions based on resonant optical cavities.

Michelson's experiment, made back in 1879 [1], is presented as one of the most relevant historical experimental proof backing the theory of Special Relativity and most particularly its first postulate which assumes all inertial systems to be formally equivalent. In view of the conceptual transcendence of the deductions made from it [2-5], it cannot be superfluous to carefully update its experimental analysis, contrasting all possible interpretative issues, and to subject it to any new insight. Since its emergence, Michelson's experiment has lead to different interpretations. Three main ones were proposed in an attempt to account for the null result of this speed-related interferometric experiment, yet preserving in great part compatibility with the classical theories then in existence. These were the aether drag assumption, the Lorentz contraction hypothesis, and the emission theories [2,3]. The aether drag hypothesis assumed that the aether frame was bound to all bodies of finite mass. The Lorentz contraction assumed that all material bodies contract by a factor of $(1 - v^2/c^2)^{1/2}$ in the direction of motion to their speed relative to the aether. In the emission theory Maxwell's equations are modified such that the speed of light remains dependent on its source own speed. However, none of these efforts to adapt classical standpoints has been thought tenable and finally a break with all these classical framed attempts was made by Einstein in 1905, introducing an interpretative framework laying in a new metric based on a speed-induced affectation of space and time and their union into a unique space-time entity conceived by Minkowsky [2-5].

In textbooks dealing with Michelson's experiment it is implicitly assumed that the beam reflected on the beam-splitter invariantly impinges back exactly at the same place after reflecting on the top mirror [2,3]. This determinant assumption cannot be taken as implicit since the interferometer is moving into space by its link to the earth's motion, and it requires thus to be carefully experimentally checked, in view of the extreme relevance of the related deductions. This decisive point has apparently been taken for granted rather than directly sounded, leaving thus under shadow any eventual small lateral shift between the recombined beams and its repercussion on the length of their effective path, and thus on its null interferometric result. In effect, depending on the incident beam being considered to exactly reflect at right angle or to slightly deviate from it due to a scanty speed-induced deflection, the paths of the transmitted and reflected beam of the interferometer have a different relationship (fig.1.a and 1.b). Such a check-up is therefore essential for any interpretation of the null result of Michelson's experiment and related ones, in view of the drastically different conceptual conclusions that may derive. A common framework for testing special relativity is the test theory of Mansouri and Sexl [6] which, in order to appraise the suitability of a given experiment for testing special relativity, conveniently employs a framework which allows its transgression. In a similar compliance, conformity with special relativity would be fulfilled in paragraph 2.1. (Reflection with deflection) and would not in 2.2. (Reflection without deflection). However, our main purpose will not center on revising interpretations derived from the null result of Michelson's experiment, but instead on checking crucial parameters determining the speed-independent performance of this interferometer. From a purely formal standpoint, the two different possibilities corresponding to the transmitted and reflected beams to coincide or not at the very same time and position on the beam-splitter, should be considered. So, let us ponder these two alternatives and their associated implications.

Let us point out that special relativity introduces also an oblique path for the transverse beam of Michelson's experiment and thus a deflection, as seen by an external observer. However, the difference lays in that within the framework of special relativity the beam deflection is an observational effect and the transverse beam has consequently as many paths as observers, while in the present framework the beam path is unique and the way different observers may see it, is just a matter of appearance.

## 2. Description and working way of the set-up to detect the presence or absence of speed-induced beam deflection

The incident beam is split by means of a thin layer beam-splitter into a reflected and a transmitted component, such as in the Michelson interferometer. Right behind the thin layer a high definition array photo-detector whose resolution can reach 5 $\mu$m [7,10], is fixed in contact to it, and so the transmitted beam impinges right on it. Alternatively, a half-transparent gold coating may be deposited directly on the array. The beam reflected on the beam-splitter, on its way back to it after reflection on the top mirror, impinges too on the detector array. The traces that the two beams leave on the array-detector are recorded and computerized.

This arrangement allows the recording of the cross-section distribution of the incident beam as well as its distribution after reflection on the top mirror. Assuming e.g. a gaussian distribution, the deconvolution of the recombined beams radial distribution would allow the differentiation between the two alternatives, according to the maintenance of a unique distribution or its splitting in two ones. Still, the beam images on the array detector may be better differentiated by using a convergent beam impinging on the beam-splitter instead of a collimated one (Fig.3.a and 3.b), since having then a different size. The transmitted beam area and cross-section distribution are first measured by the high-resolution array-detector, with the reflected beam path being blocked by means of a shutter. Afterwards, the path of the reflected beam is unlocked, and in being convergent it impinges back on the beam-splitter as a spot, which superposes to the wider area of the incident beam. Due to the smallness of the deflection angle at ordinary speeds, the mechanical angular resolution of the beam-splitter and top-mirror should be of the order of second of arc. However, there is a way to skip such a strict requirement by centering the trace of the beam when first impinging on the beam-splitter with its trace upon reflection on the top mirror. To be fulfilled, the equivalency of inertial systems implies that the two beam spots on the array-detector should remain immobile in changing the system orientation. They should also remain steady under the variation of the speed module.

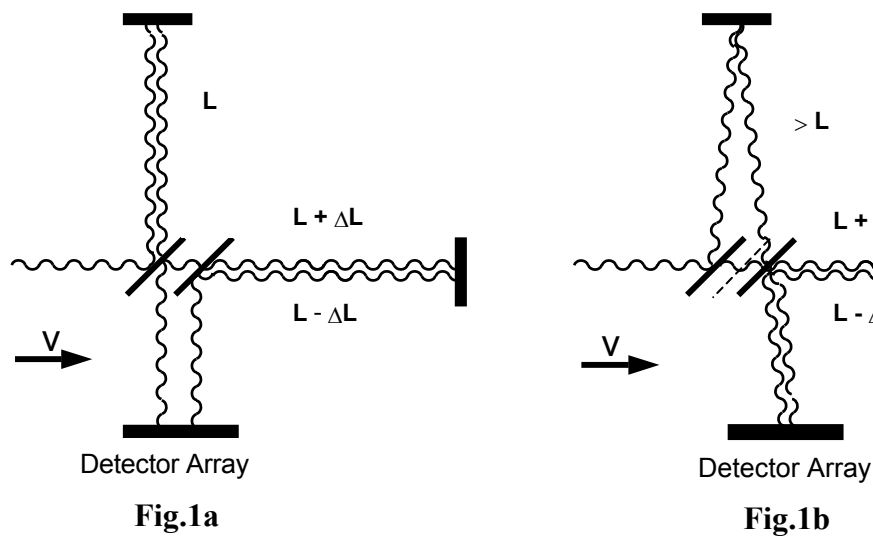

**Fig.1.a and 1.b:** *Deflection and deflectionless alternatives in Michelson´s experiment*

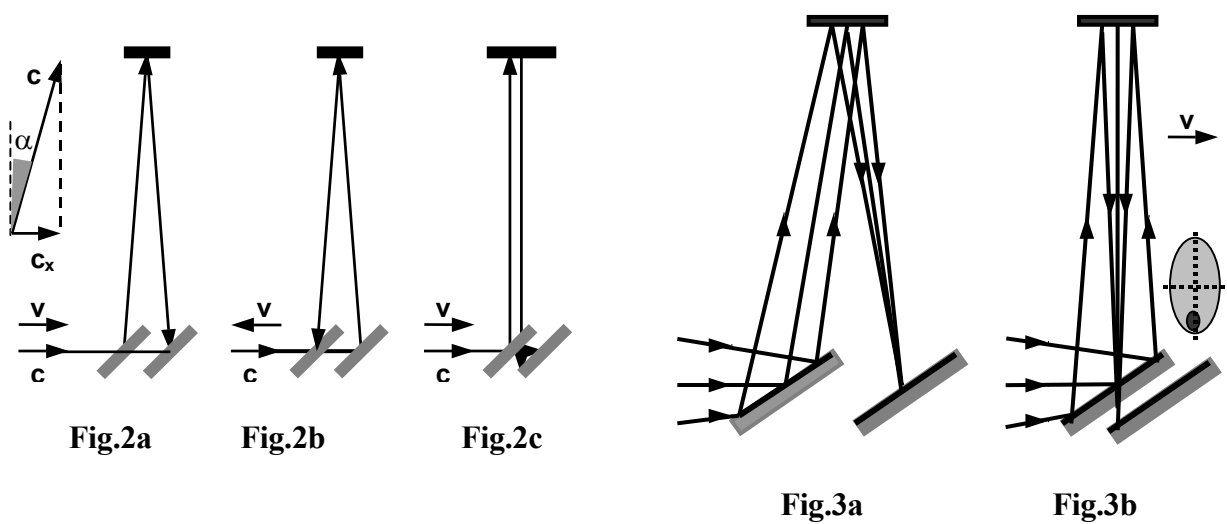

**Fig.2a** and **2b. Reflection with deflection:** *The reflected beam impinges back exactly at the same point. This implies that it has been slightly deflected and hence its Y up-down path is slightly longer than the double of the distance d between beam-splitter and mirror. So, with deflection the path length up to the X-axis is: Y = 2 (d + ∆d), where ∆d=d /2 ($v^2/c^2$)*

**Fig.2c. Reflection without deflection:** *The reflected beam is not deflected and so it does not impinge back at the same point. So, without deflection the path length of the Y beam up to the X-axis is: Y = 2 d. Its length up to the beam-splitter is shorter or longer according to the relative direction of the beam and the system speed*

**Fig.3a. Convergent beam with deflection:** *Epicentric traces of the incident beam and once impinging back on the detector array. This case implies inertial frames to be equivalent at first order dependence on (v / c)*

**Fig.3b. Convergent beam without deflection:** *Off-centric traces of the incident beam and once reflected back. In that case inertial frames would not be equivalent at first order (v / c)*

Thus, the device plays simultaneously two functions, that of a beam-splitter and that of a high definition array detector, providing at once three important beam parameters by measuring the spot size, its cross-section distribution and its position on the array. There are crucial data to ascertain the interpretation of the underlying cause of the interferometer null result, allowing to discriminate between the two alternatives considered in order to achieve a complete analysis.

### 2.1. *Reflection with Deflection*

For simplify, let us select the case in which the interferometer is moving in the same direction as the light beam (fig.2.a and 3.a). For first-order equivalency to be fulfilled the reflected beam must invariantly coincide precisely at the same position in its way up and down, and hence the component $c_x$ of the speed of light along the direction x of the interferometer motion must necessarily be equal to the system speed v.

$$c_x = v \qquad \text{where:} \qquad c_x = c \cdot \sin \alpha$$

This implies that the beam is deflected by an angle equal to:

$$\alpha = \arcsin (c_x / c) = \arcsin (v / c)$$

The deflection has always the direction of the system motion. For example, if the speed v points to the right and the speeds $c_x$ and v are parallel ($c_\rightarrow v_\rightarrow$) then the rotation of $\alpha$ is clockwise (Fig.2.a), and if $c_x$ and v are anti-parallel ($_\leftarrow c\ v_\rightarrow$) it is anticlockwise (Fig.2.b). In both cases the beam impinges at the same point on the beam-splitter, since the deflection always goes in the direction of motion and exactly counterbalance the beam-splitter dislodge. This requires it to reflect at 90º ± $\alpha$ according to the relative speed directions of the beam and the beam-splitter. So, the reflected beam will impinge back at the same position as the incident beam only if the speed component $c_x$ of light along the system motion direction is equal to the system speed v, since in that case there is no relative speed between them. Consequently, the lateral speed of the beam is equal to the system speed and thus their motion along the axis is synchrone, and from the point of view of the moving system there is no beam deflection, nor shift when impinging back on the beam-splitter.

**(a)** Let us stress that the deflection angle is assumed to be independent of the beam wavelength used, and so is it for the result of Michelson's experiment.

**(b)** The effective deflection angle $\alpha$ of the beam path is not fixed by the relative velocity between the system and any eventual observer, but exclusively by the relative velocity between the beam-splitter and light. This relative velocity fixes the impact energy of photons on the mirror, which in turn fixes the deflection angle.

**(c)** The definition of this angle is single, since it is independent of the source speed but only fixed by the relative velocity between the system and light, in contrast with its definition within the context of special relativity, in which there are as many deflection angles as observers. In the present framework the beam trajectory within space is unique and it corresponds to the viewpoint of the impinging photons, which is single, and not to that of observers, which leads to a multiplicity of deflection angles and beam paths.

**(d)** Which could be the physical cause of the deflection angle? It appears as a function of the ratio v / c, however this relation may be regarded as the simplified result of the ratio pc / mc². Hence:

$$\sin \alpha = E_r / E_i = pc / mc^2 = mvc / mc^2 = v / c$$

where $E_r = pc = h\nu$ and $E_i = mc^2 = h\nu_0$. $E_r$ stands for the relative collision energy, i.e. that measured by a detector on the moving system, and $E_i$ for the intrinsic collision energy, i.e. that a detector at absolute rest would measure [9,13,14], $\nu$ is the frequency perceived by the inertial system and $\nu_0$ the absolute frequency. Here what is regarded as absolute rest is the absence of motion relative to the Cosmic Microwave Background (CMB) in the same way as absolute speed is defined for mechanical waves, e.g. such as the air for sound. In fact, any field can be considered to be at rest when taken as a whole, since their carrier speed is constant and equal to that of light. Consequently, fields cannot have a drift velocity and thus the electromagnetic field may gainfully be used as an extended absolute reference.

The deflection angle can be looked at as determined by the ratio of the collision energy of the impinging photons against the beam-splitter and the intrinsic energy of the impinging photons. The collision strength determines the magnitude of the deflection [8,9] and may be defined as:

$S = k (v / c) = k (pc / mc^2) = k (mvc / mc^2) = k (h\nu / h\nu_0)$

Equivalently, the beam deflection would be the consequence of a tangential momentum of the reflected photons, i.e. in the direction of the beam-splitter motion (fig.1.b, 2.a and 3.a). The same applies to photons from the light source itself [9].

**(e)** In order to impinge back at the same position on the beam-splitter, the deflection angle must be equal to: $\alpha$ = arc sin (v / c). Its formulation is similar to Bradley aberration of light: $\alpha$ = arc tg (v / c), even though their underlying physical processes are different, since respectively due to a deflection (sin $\alpha$ = $E_r$ / $E_i$ = mvc / mc$^2$ = v / c) and to a time of flight (tg $\alpha$ = $L_x$ / $L_y$ = v $\Delta$t / c $\Delta$t = v / c). A suggestive coincidence.

**(f)** The speed-induced beam deflection would keep constant the spot position on the beam-splitter.

## 2.2. *Reflection without Deflection*

This second alternative assumes that the beam reflects perpendicularly to itself whatever the system motion, or in other words, independently of the beam-splitter speed and therefore independently of the beam collision strength on it. In that case the beam does not impinge back at the same point since the beam-splitter has meanwhile moved away, laterally to the beam path, while it was going up and down, apart from the system motion. So, in the case of no speed-induced deflection the beam would impinge back slightly shifted from the initial impinging point on the beam-splitter (Fig.2.c and 3.b). Consequently, since the beam-splitter is inclined at 45º, the beam impinges on it at a lower or upper position than its initial one, depending on both speeds c and v having the same or opposite orientation, respectively.

Let us define the order of magnitude of the effect to be measured, i.e. that of the speed-induced lateral drift of the beam in absence of deflection. In that case, while the reflected beam reaches back the beam-splitter, the latter would have drifted a distance:

$2 \Delta x = v \cdot t = v (2d / c) = 2 d (v / c)$

Thus the length the reflected beam (Y) must cover to reach back the beam-splitter at the same position is:

$d(Y) = 2 (d^2 + \Delta x^2)^{1/2}$

So, the reflected beam (Y) must cover a path excess $\Delta L$ to the distance between beam-splitter and top mirror:

$\Delta L = 2 d(Y) - 2 d = 2 (d^2 + \Delta x^2)^{1/2} - 2 d = 2 (d^2 + d^2 \cdot v^2 / c^2)^{1/2} - 2 d = 2 d (1 + v^2 / c^2)^{1/2} - 2 d$

$\Delta L \cong 2 d (1 + 1/2 \cdot v^2 / c^2) - 2 d = 2 d (1/2 \cdot v^2 / c^2) = d (v^2 / c^2)$

For d = 1 m and the earth orbital speed v $\approx$ 30 km/s, 2 $\Delta x$ = 2 (30 / 300000) = 2.10$^{-4}$ m $\approx$ 200 $\mu$m and 2$\Delta$L = 2.10$^{-8}$ m $\approx$ 0.02 $\mu$m.

Let us point out that the lateral drift $\Delta x$ has a first order dependence (v / c), while the longitudinal path stretching $\Delta L$ relative to the distance d between beam-splitter and top mirror has a second order dependence (v$^2$ / c$^2$). A lateral drift of the order of magnitude calculated would be easily detected by the high resolution array, but since it should be expected that such a shift would have been detected by earlier analogue experiments, it can be anticipated that the alternative of deflection absence can hardly be retained.

In order to get optimized performances, some critical parameters of the set-up are the beam shape, divergence, beam pointing stability, and optics surface quality, but we think that the present state of the art would allow to achieve easily the required resolution to back first-order equivalency. Also, the computer-treatment of the digitalized image of the spots on the high-resolution array-detector would allow to further improve experimental efficiency.

## 3. Discussion

Let us highlight that the speed-induced deflection angle $\alpha$ is not an observational effect, in the sense that it does not correpond to the relative perception of an arbitrary observer, as it is in special relativity in which there are as many deflection angles as inertial observers. It is instead an actual deflection, whose angle $\alpha$ is absolute since fixed by the absolute speed v of the inertial system. So, the standard definition from the theory of relativity of the system speed v which is always relative has been avoided in view that this standpoint leads to as many beam paths and deflection angles as observers, whose multiplicity necessarily implies all these paths to be only apparent. The beam path into space is logically unique since a unique beam cannot have at once different paths, and the many ways observers may see it are therefore not relevant in the present analysis, which is only concerned with the intrinsic path traced by the beam in its journey through space, independently of the point of view of any observer. However, let us highlight that the proficiency of Minkowski metric and special relativity is not queried but only are specific interpretative aspects of their physical meaning. The fact that the specific equivalency of inertial systems considered here has been attributed to a speed-induced deflection of light rays instead of to a speed-induced alteration of space and time as in special relativity, does not imply a mutual exclusion. In effect, it should be avoided to mix the properties of the metric with those of physical reality itself. The map should never be confounded with the territory itself, in other words the theoretical mathematical construct should never be assimilated to the proper physical reality. Different conceptualizations may provide different grades of access to the underlying reality.

However, the speed-induced deflection angle $\alpha$ cannot be covered by the fundamentals of special relativity since it denies the possibility of any absolute reference, and keeps denying it in spite of the latter discovery of the Cosmic Microwave Background (CMB). This is a self-imposed limitation which restrains its application range and impedes an exhaustive access to physical reality. For example, the actual deflection angle in Michelson experiment is not dependent on the relative velocity between the light source and the observer, but dependent instead on the system absolute velocity. Let us mention that absolute speed has already been defined relative to the CMB. In effect, for practical means the CMB constitutes a suitable absolute reference since being highly isotropic, shaped according to a perfect blackbody spectrum, and of universal range. It has already been established that our solar system drifts at an absolute speed of about 370 km/s towards the Leo constellation.

In view of the postulated denial in 1905 of any absolute reference, sixty years before the discovery of the CMB in 1965, and based on the then recent aether refutal, special relativity still wants to regard the CMB just as a mere "preferred" referential frame, not affecting in any way its grounds. However, this is a doctrinal standpoint neglecting the fundamental fact that the CMB is a unique *non-local* reference. In other words, the CMB constitutes a unique *extended* referential frame which covers the whole universe. The CMB differs fundamentally from local references used by special relativity, all being inertial systems, since it is neither an inertial nor a local system. To disregard these facts leads to cut down physical reality. Furthermore, it allows to *normalize* the speed of all inertial systems and can be used just as the presumed aether was intended to. In fact, since there are $10^9$ photons per massive particle, mostly stars hydrogen, the usual conceptual perspective can be inverted and the material part of the universe can be seen as a disperse condensed residue, immerse in an electromagnetic "atmosphere". The effects of motion on material and thus finite inertial systems can hence be advantageously referred to this non-inertial, non-local and non-finite but instead wholly extended, steady system.

Nevertheless, absolute speed does not necessarily need a medium to be referred to, as commonly thought. Absolute speed can be conceptually defined just from the intrinsic properties of light, and so in the definition of the deflection angle $\alpha$ *the absolute reference to which speeds are referred to is light itself*, through the path of light rays taken as a reference. Furthermore, the deflection angle has been correlated to the ratio between the beam energy perceived by the moving system and the corresponding one if it would be at rest into space. Let us stress that this standpoint on absolute speed does not appeal to any interpretation of the nature of space. Our main concern has been to experimentally relate by means of the original set-up proposed the optical equivalency of all inertial systems to the unique path of the reflected beam, which would be deflected according to the absolute speed of the reflecting element. This is equivalent to consider the relationship between light and the system motion, or more concretely, between the component $c_x$ of the speed of light along the direction of the beam-splitter motion and its own speed v. For the reflected beam to always impinge back at the same position on the beam-splitter whatever its speed v, the deflection angle must be such that $c_x = v$. In other words, for the speed-induced deflection to lead to the same result on any inertial system, the speeds $c_x$ and v must always be equal. This is an imperative requirement for inertial systems to be equivalent, on first-order, in regard to light beams.

Moreover, the fact that special relativity postulates that all inertial systems are equivalent, implies taking for granted that not a single physical effect could be dependent on the system absolute speed into space, concept that does not even make sense within its framework. Nevertheless, an amazing point arises from the fact that what would make all inertial frames to be equivalent, at least on first-order (v/c), depends just on their absolute speed into space. That is to say, the validity of special relativity would rely on a magnitude that it denies, since what makes all inertial frames to be first-order equivalent would derive from the speed-induced deflection of light, which ironically is determined by the system absolute speed, since the speed v is referred to c through its component $c_x$ along the direction of motion (v = $c_x$ = c .sin $\alpha$). But, as the speed of light is constant in vacuum it constitutes an absolute reference by excellence, and since the value of v is related with c, it expresses thus an absolute speed, which can be conveniently defined through the CMB. Let us mention that finding a way to avoid or to act upon the presumed speed-induced deflection of light would allow to detect the own speed of inertial systems [7,9,13].

We have proposed a causal explanation for the first-order optical equivalency of inertial systems, at least of the specific one concerned with the path of light beams. The assumption of speed-induced deflection brings a causal grasp to the specific equivalency of inertial systems since it allows the invariance of the spot position on the array detector of the reflected beam and its coincidence with the direct beam spot. We have replaced the speed-induced alteration of space and time of special relativity by the speed-induced deflection of light rays. The underlying cause for all physical effects to apparently behave independently of the system motion, or so to say, for inertial systems to appear exhaustively equivalent, might so be better understood. It would be worthwhile to keep deepening the causal understanding of the pertinence of the postulated equivalency of all inertial systems, and to sound experimentally some epistemological aspects of special relativity in an effort to deepen and widen our knowledge beyond postulated grounds up to causal foundations. It should be avoided to be definitively "formatted" by any theory, and carefulness should prevail when making a transcription of any mathematical description to a physical one. The worst enemy of scientific progress is the progressive settlement of a theory as a dogma.

## 4. Conclusion

The experimental set-up proposed would allow to define a type of optical equivalency based on ray optics and specific to the effect of speed on the reflection of light beams and of first order dependence on v. In a similar way as the null result of Michelson experiment has endorsed the emergence of a specific metric assuming space contraction and time dilatation, the invariance of the spot position on the high-resolution photo-detector array, acting also as beam-splitter, would back a specific equivalency of inertial system based on the speed-induced deflection of light. As far as we know it is the only design that would detect any deviation from first-order optical equivalency through the detection of any speed-induced shift between the trace of an impinging beam on a high-resolution array-detector and its trace on impinging back to it after reflection on a top mirror. Let us point out that the beam perceives the velocity of the inertial system exclusively in the instant of interaction with its components, i.e. beam-splitter, mirror and detector. In vacuum, during the lapse of time the beam is travelling in between them, it is not interacting with the inertial system and thus it cannot be affected by its velocity. Both are moving independently as long as the refractive index is equal to one, and therefore there is no Fresnel partial drag of the light beam [2,3]. In the hypothetical deflectionless case the set-up would have a first order (v/c) sensitivity to its own speed through space, but in the case of deflection it is blind to it. In assuming a priori the equivalency of inertial systems it is expected the experiment to prove the two beam traces on the array-detector to remain steady independently of the orientation of the speed vector and of its module, providing thus a novel experimental proof of the optical equivalency of inertial systems on the first order.

Let us point out a consequence of the equivalency of inertial systems. The failure so far to measure absolute speed of inertial systems through the exclusive use of local light sources would derive from their equivalency, and this leads to a vicious circle. In effect, how could we expect apparatus not to provide systematically the same value for the speed of light relative to themselves if they are unable to detect their own speed? How could they sum up the speed of light to a magnitude they do not detect? So, it is proper to consider invariant the speed of light but it is also tenable to wonder about the intrinsic or extrinsic nature of the invariance of the speed of light. Couldn't we be systematically measuring the intrinsic speed of light through space, since optical set-up are unable to detect their own speed? The article aims to highlight that the invariance of the speed of light may not necessarily be a formal property, in other words it could be the consequence of experimental shortcuts. The difference between these two issues is extreme, since in the first case it would represent an intrinsic property of nature whereas in the second case it would only be an extrinsic property rising from experimental limitations in the detection of specific properties, leading thus to an incomplete access to physical reality. We should be careful not to attribute to nature ultimate properties that could only derive from our present experimental state of the art and might not stand technical breakthroughs. The device proposed has the collateral purpose of helping unravel why apparatus do not detect their own speed through space, and at the same time sounding if the invariance of the light speed is an intrinsic or extrinsic property. We have proposed an original set-up to evidence a specific equivalency of inertial systems and have based the related analysis on causal grounds, founded on the speed-induced deflection of light rays. Let us suggest that searching for an access to the parameters governing the speed-induced light rays deflection could allow envisioning to reach self-speed detection, e.g. looking for an eventual way to cancel out or alter the speed-induced beam deflection. Hence, we would like to bring our contribution to the understanding of the cause of experimental failure in the detection of the proper speed, and despite odds, to encourage the retaking of the search for self-speed sensitive devices, making the most of current technological possibilities.

*Author's other related articles:*

First and second order electromagnetic equivalency of inertial systems, based on the wavelength and the period as speed-dependant units of length and time.

A dual set-up based on Bradley's aberration of light, using simultaneously stellar and local light sources.

lanl.arXiv.org e-Print archive (url: xxx.lanl.gov),  Physics, Subj-class: General Physics